\documentclass[%
 reprint,
superscriptaddress,
 amsmath,amssymb,
 aps,
]{revtex4-2}

\usepackage{graphicx}
\usepackage{dcolumn}
\usepackage{bm}

\usepackage{physics}
\usepackage{mhchem}
\usepackage{natbib}

\usepackage{color}

\usepackage[normalem]{ulem}

\begin{document}

\title{Equation of state of spin-polarized nuclear matter in the relativistic Hartree-Fock method}

\author{Toi Tachibana}
\affiliation{Department of Physics, Kyoto University, Kyoto 606-8503, Japan}

\author{Kouichi Hagino}

\affiliation{Department of Physics, Kyoto University, Kyoto 606-8503, Japan}
\affiliation{Institute for Liberal Arts and Sciences, Kyoto University, Kyoto 606-8501, Japan}
\affiliation{ 
RIKEN Nishina Center for Accelerator-based Science, RIKEN, Wako 351-0198, Japan}

\author{Kenichi Yoshida}
\affiliation{Research Center for Nuclear Physics, The University of Osaka, Ibaraki, 567-0047, Japan}
\affiliation{ 
RIKEN Nishina Center for Accelerator-based Science, RIKEN, Wako 351-0198, Japan}
\affiliation{Center for Computational Sciences, University of Tsukuba, Tsukuba, Ibaraki 305-8577, Japan}

\author{Qiang Zhao}
\affiliation{China Institute of Atomic Energy (CIAE), P. O. Box 275(18), Beijing 102413, China}
\affiliation{Research Center for Nuclear Physics, The University of Osaka, Ibaraki, 567-0047, Japan}

\date{\today}

\begin{abstract}
We calculate 
the equation of state (EOS) of spin-polarized nuclear matter in the relativistic Hartree-Fock method. 
To this end, we employ the relativistic point-coupling model, with which 
the Fock terms are considerably simplified, reducing them to the same form as the Hartree terms.
In analogy to the slope parameter $L$ of the isospin-symmetry energy for spin-unpolarized matter, 
we evaluate the spin slope parameter $L_s$ of the corresponding spin-symmetry energy 
for spin-polarized matter. We find that 
the slope parameter $L$ and the spin slope parameter $L_s$ 
have a negative correlation in the case of isoscalar polarization, where neutrons and protons 
are spin-polarized in the same direction.
On the other hand, 
the spin slope parameter is nearly independent of the slope parameter 
in the case of isovector polarization, where neutrons are spin-polarized along the opposite direction to protons. 
We show that these correlations are  
a natural consequence of the relativistic point coupling model which we employ.
\end{abstract}

\maketitle


\section{INTRODUCTION}
\label{sec:intro}

The equation of state (EOS) of nuclear matter 
plays a crucial role 
in a variety of astrophysical phenomena, such as the inner structure of neutron stars and supernovae \cite{Yamamoto_Togashi,Togashi_2017,Shen,Hempel_2012,Sumiyoshi_2005}. 
It has been regarded as one of the major goals of nuclear physics to  
determine the EOS of nuclear matter with high precision, 
using a unified method both for finite nuclei and infinite nuclear matter. 

The EOS of symmetric nuclear matter in the vicinity of the saturation density 
is characterized by the saturation density, the saturation energy and 
the incompressibility, which have 
already been known to some extent \cite{RingSchuck,Bohr69,Sagawa21,FetterWalecka,K_value-2,K_value-3,GARG201855}.
However, this information is not sufficient to discuss 
the inner structure of neutron stars, as a high isospin asymmetry and a wide range of nucleon density 
are realized there.
Therefore, many theoretical and experimental studies have been carried out \cite{APR,Togashi_2013,Bompaci_1991,Lattimer_2013,Berman_1975,Zenihiro,PREX,CREX,Tsang2012,Carbone2010} 
aiming at extending the EOS into the isospin-asymmetric and a wide density regions. 

In particular, it has been expected that 
the symmetry energy near the saturation density can be constrained by terrestrial nuclear experiments.
Here, the symmetry energy approximately corresponds to 
the difference between the energies of symmetric nuclear matter and pure neutron matter,  
characterizing 
the EOS of asymmetric nuclear matter. 
For instance, it is known that the slope parameter defined as the derivative of the symmetry energy at the saturation 
density is linearly correlated with 
the neutron skin thickness of heavy nuclei such as \ce{^{208}Pb} \cite{Roca-maza}, and thus 
measuring the neutron skin thickness of finite nuclei can constrain the slope parameter. Because of this, 
numerous experiments 
to measure the neutron skin thickness have been carried out so far \cite{Zenihiro,Tamii2011,PREX,CREX}.

While 
spin saturated nuclear matter is usually considered in such studies,  
the EOS of symmetric nuclear matter 
can be extended not only to the isospin-asymmetric direction, that is, neutron matter, 
but also 
to the spin-asymmetric direction, that is, spin polarized matter. 
This is to 
explore properties of nuclear matter in the two-dimensional 
spin-isospin space.  
Spin-polarized matter is of interest for several reasons.
Firstly, 
the EOS of spin-polarized matter determines 
the existence of a ferromagnetic phase of nuclear matter 
through its 
magnetic susceptibility \cite{Vidana_2002,Niembro_1990,Niembro_1991,Maruyama}.
Secondly, 
spin polarization in nuclear matter may affect the proton fraction in neutron star matter, 
which may enhance neutron star cooling \cite{Khoa_2020}.
Recently, observations of the blue kilonova ejecta following the neutron star merger GW170817 \cite{kilonova,GW170817,Evans} 
have indicated that a hypermassive neutron star remnant has a strong magnetic field of $(1-3)\times10^{14}$ G at its surface \cite{Metzger_2018}, 
and a partial spin polarization could occur there.
Because of this, the EOS of spin-polarized matter has attracted lots of attention in recent years.
In Refs. \cite{Khoa_2022,Khoa_2020}, 
Khoa et al. discussed the EOS of spin-polarized matter using the non-relativistic Hartree-Fock method with the CDM3Yn 
interaction and concluded that up to 60-80\% of baryons in the neutron star merger may be spin polarized. 
They also defined the spin slope parameter as the derivative of the EOS of fully spin-polarized matter 
at the saturation density. 
Properties of spin polarized nuclear matter were studied also in Ref. \cite{Margueron2009,Chamel2010} using the non-relativistic 
Skyrme interactions. 

In this paper, we 
study the spin-polarized nuclear matter
by using the relativistic framework, 
which more naturally treats the spin degree of freedom of nucleons than non-relativistic frameworks. 
In addition, 
we investigate the correlation between the slope parameter and the spin slope parameter.
If there is a significant correlation between these slopes, 
one can indirectly constrain the EOS for spin-polarized matter by measuring the neutron skin thickness or other observables 
which are correlated with the slope parameter. Notice that there have been almost no discussions so far on experimental 
probes which can constrain the EOS of spin-polarized matter. 

To discuss the EOS of spin polarized matter, 
it has been pointed out that the Fock exchange terms 
play a crucial role \cite{Niembro_1990,Maruyama}. 
For this reason, we shall 
adopt the relativistic point-coupling model, which simplifies the treatment of 
the Fock terms. 

The paper is organized as follows.  
In Sec. \ref{se:formalism}, we formulate 
the relativistic point-coupling model. 
In Sec. \ref{sec:results}, we present the numerical results for the spin polarized EOS 
for several values of the spin polarization rate. 
In Sec. \ref{sec:correlation}, we discuss the correlation between the slope parameter and the spin slope parameter,  
and propose a possible probe
to constrain the EOS for spin-polarized matter. 
Finally, we summarize the paper in Sec. \ref{sec:summary}.


\section{Relativistic Point-Coupling Model}
\label{se:formalism}

In order to calculate the EOS of spin-polarized matter, we employ the relativistic 
zero-range point-coupling model \cite{PC-F1,PC-PK1}. This model was originally invented for 
Hartree calculations of atomic nuclei, but it has recently been extended to Hartree-Fock 
calculations with the exchange terms \cite{Sulaksono_2003,Liang_2012,PCF-PK1}. 
The Lagrangian for this model reads 
\begin{align}
    \label{eq:Lag}
    \mathcal{L} = \mathcal{L}^{\text{free}} + \mathcal{L}^{\text{4f}}
        + \mathcal{L}^{\text{der}} + \mathcal{L}^{\text{em}},
\end{align}
where $\mathcal{L}^{\text{free}}$ is the free nucleon term given by 
\begin{align}
    \mathcal{L}^{\text{free}}=\bar{\psi}(i\gamma_{\mu}\partial^{\mu}-M)\psi. 
\end{align}
Here, $\psi$ and $M$ represent the nucleon field and the nucleon mass, respectively. 
$\mathcal{L}^{\text{4f}}$ refers to the four fermion term, and is 
expressed as
\begin{align}
    \label{eq:Lag_4f}
    \mathcal{L}^{\text{4f}} = 
        & - \frac{1}{2} \alpha^{(0)}_S (\bar{\psi} \psi)^2
        - \frac{1}{2} \alpha^{(0)}_{tS} (\bar{\psi} \vec{\tau} \psi)^2 \notag \\
        & - \frac{1}{2} \alpha^{(0)}_V (\bar{\psi} \gamma^{\mu} \psi)^2
        - \frac{1}{2} \alpha^{(0)}_{tV} (\bar{\psi} \gamma^{\mu} \vec{\tau} \psi)^2 \notag \\
        & - \frac{1}{2} \alpha^{(0)}_{PV} (\bar{\psi} \gamma_5 \gamma^{\mu} \psi)^2
        - \frac{1}{2} \alpha^{(0)}_{tPV} (\bar{\psi} \gamma_5 \gamma^{\mu} \vec{\tau} \psi)^2 \notag \\
        & - \frac{1}{2} \alpha^{(0)}_{PS} (\bar{\psi} \gamma_5 \psi)^2
        - \frac{1}{2} \alpha^{(0)}_{tPS} (\bar{\psi} \gamma_5 \vec{\tau} \psi)^2 \notag \\
        & - \frac{1}{2} \alpha^{(0)}_T (\bar{\psi} \sigma^{\mu\nu} \psi)^2
        - \frac{1}{2} \alpha^{(0)}_{tT} (\bar{\psi} \sigma^{\mu\nu} \vec{\tau} \psi)^2.
\end{align}
Here, the subscript $S$, $V$, $PS$, $PV$ and $T$ represent the scalar, vector, pseudoscalar, pseudovector and tensor coupling terms, respectively, while the additional subscript \lq\lq$t$" refers to the isovector
channel.
For uniform matter, the contribution from the pseudoscalar term vanishes due to parity conservation. 
Likewise, the derivative coupling term $\mathcal{L}^{\text{der}}$ and the electromagnetic term $\mathcal{L}^{\text{em}}$ 
in Eq. \eqref{eq:Lag} 
do not contribute to the EOS of infinite nuclear matter.

With the no-sea approximation, one can expand the nucleon field $\psi(x)$ 
for uniform matter 
by using the basis of annihilation operators of nucleons $\{c_{\bm{p}st}\}$ together with 
the positive energy plane wave solutions $u(\bm{p},s,t)$ as, 
\begin{align}
    \label{eq:expansion}
    \psi(x) = \frac{1}{\sqrt{V}} \sum_{\bm{p},s,t} e^{-ip \cdot x} u(\bm{p},s,t) c_{\bm{p}st},
\end{align}
where $\bm{p}$ represents the single-particle momentum and the indices $s$ and $t$ refer to the spin and the isospin of a nucleon, respectively. 
$V$ represents the volume of the system and we take the infinite volume limit in the following discussion.
In the Hatree-Fock approximation, the ground state wave function is given 
as a Slater determinant,
\begin{align}
    \label{eq:Slater_det}
    \ket{\Phi_0} = \prod_{\bm{p},s,t} c^{\dagger}_{\bm{p}st} \ket{-},
\end{align}
where $|-\rangle$ is the vacuum state. 
Taking the expectation value of the corresponding Hamiltonian in the ground state yields 
the energy density of the system,
\begin{align}
    \label{eq:energy_density}
    \mathcal{E} = \frac{\bra{\Phi_0}H\ket{\Phi_0} }{V}
    = \mathcal{E}_{\text{kin}} + \mathcal{E}_{\text{4f}}.
\end{align}
In this equation, 
the kinetic energy term $\mathcal{E}_{\text{kin}}$ is expressed as
\begin{align}
    \label{eq:kin_energy_density}
    \mathcal{E}_{\text{kin}} = \sum_{s,t} \int \frac{d^3p}{(2\pi)^3} 
        u^{\dagger}(\bm{p},s,t) (\bm{p}\cdot\bm{\alpha}+M\beta)u(\bm{p},s,t),
\end{align}
where $\bm{\alpha}=\gamma^0\bm{\gamma}$ and $\beta=\gamma^0$ are Dirac matrices. 
The four-fermion interaction energy $\mathcal{E}_{\text{4f}}$ is written as a sum of the 
Hartree (direct) and the Fock (exchange) terms, 
\begin{align}
    \mathcal{E}_{\text{4f}} = \mathcal{E}^{\text{D}}+\mathcal{E}^{\text{EX}},
\end{align}
with 
\begin{align}
    \mathcal{E}^{\text{D}}&=\frac{1}{2}\sum_{i}\alpha^{(0)}_i\left[\sum_{\xi} \bar{u}(\xi)\left(\mathcal{O}\Gamma\right)_iu(\xi)\right]^2, 
    \label{eq:D}\\
    \mathcal{E}^{\text{EX}}&=-\frac{1}{2}\sum_{i}\alpha^{(0)}_i \sum_{\xi,\xi'}
    \left[\bar{u}(\xi)\left(\mathcal{O}\Gamma\right)_i u(\xi')\right] 
    \left[\bar{u}(\xi')\left(\mathcal{O}\Gamma\right)_i u(\xi)\right]. \label{eq:EX}
\end{align}
Here, we have used a simplified notation, in which $\xi$ represents a set of $(\bm{p},s,t)$ and the sum $\sum_{\xi}$ is defined as
\begin{equation}
    \sum_{\xi}=\sum_{s,t}\int\frac{d^3p}{(2\pi)^3}. 
\end{equation}
In Eq. \eqref{eq:D} and \eqref{eq:EX}, 
the matrices $\mathcal{O}$ and $\Gamma$ represent the isospin matrix and the gamma matrix, respectively, with 
\begin{align}
    \mathcal{O}\in\{1,\tau_3\}, ~~~~
    \Gamma\in\{1,\gamma_5,\gamma^{\mu},\gamma_5\gamma^{\mu},\sigma^{\mu\nu}\}.
\end{align}
Each combination of $\mathcal{O}$ and $\Gamma$ represents 
a coupling channel, $i\in\{S, tS, V, tV, PS, tPS, PV, tPV, T, tT\}$, that is, $\left(\mathcal{O}\Gamma\right)_i$. 
The Fierz transformation simplifies the Fock terms into
\begin{align}
    \mathcal{E}^{\text{EX}}&=-\frac{1}{2}\sum_{i,j}\alpha^{(0)}_i\Lambda_{ij}\left[\sum_{\xi} \bar{u}(\xi)\left(\mathcal{O}\Gamma\right)_ju(\xi)\right]^2,
\end{align}
where $\Lambda$ refers to the Fierz transformation matrix.
Therefore, one can easily sum up the Hartree and the Fock terms and obtains 
\begin{align}
    \mathcal{E}^{\text{4f}}&=\frac{1}{2}\sum_{i,j}\alpha^{(0)}_i(\delta_{ij}-\Lambda_{ij})\left[\sum_{\xi} \bar{u}(\xi)\left(\mathcal{O}\Gamma\right)_ju(\xi)\right]^2 \\
    &\equiv 
    \frac{1}{2}\sum_{j}\alpha_j\left[\sum_{\xi} \bar{u}(\xi)\left(\mathcal{O}\Gamma\right)_ju(\xi)\right]^2,
\end{align}
with $\alpha_j\equiv\sum_i\alpha^{(0)}_i(\delta_{ij}-\Lambda_{ij})$. 
As a result, one obtains the four-fermion interaction energy $\mathcal{E}^{\text{4f}}$ composed of 
the channel densities as, 
\begin{align}
    \label{eq:4f_energy_density}
    \mathcal{E}^{\text{4f}} & = 
    \frac{1}{2} \alpha_S \rho_S^2 + \frac{1}{2} \alpha_{tS} \rho_{tS}^2
    + \frac{1}{2} \alpha_V \rho_V^2 + \frac{1}{2} \alpha_{tV} \rho_{tV}^2 \notag \\
    & - \frac{1}{2} \alpha_{PV} \rho_{PV}^2 - \frac{1}{2} \alpha_{tPV} \rho_{tPV}^2
    + \alpha_T \rho_T^2 + \alpha_{tT} \rho_{tT}^2,
\end{align}
with 
\begin{subequations}
    \begin{align}
        \rho_S & = \sum_{\xi} \bar{u}(\xi) u(\xi),~~~\rho_{tS} = \sum_{\xi} \tau_3(\xi)\bar{u}(\xi) u(\xi), \\
        \rho_V & = \sum_{\xi}u^{\dagger}(\xi) u(\xi),~~~\rho_{tV} = \sum_{\xi}\tau_3(\xi)u^{\dagger}(\xi) u(\xi),\\
        \rho_{PV} & = \sum_{\xi}u^{\dagger}(\xi)\Sigma_z u(\xi),~~~\rho_{tPV} = \sum_{\xi}\tau_3(\xi)u^{\dagger}(\xi)\Sigma_z u(\xi),\\
        \rho_T & = \sum_{\xi}u^{\dagger}(\xi)\beta\Sigma_z u(\xi),~~~\rho_{tT}  = \sum_{\xi}\tau_3(\xi)u^{\dagger}(\xi)\beta\Sigma_z u(\xi).
    \end{align}
\end{subequations}
Here, $\Sigma_z$ is the spin operator along the $z$-axis and is expressed as 
\begin{align}
    \Sigma_z = \left(
    \begin{array}{cc}
         \sigma_z&0 \\
         0&\sigma_z 
    \end{array}
    \right)
\end{align}
in the Dirac representation.
Notice that not all of the coupling constants in Eq. \eqref{eq:4f_energy_density} are independent, 
since the rank of $1-\Lambda^{T}$ is 5,
that is smaller than the dimension of the matrix. 
The number of independent parameters in Eq. \eqref{eq:4f_energy_density} is 5, that is 
the same as the rank of the matrix. 
In Ref. \cite{PCF-PK1}, those 5 independent parameters are taken to be 
$\alpha_S, \alpha_V, \alpha_{tS}, \alpha_{tV}$ and $\alpha_T$, and the other coupling constants are expressed as 
a linear combination of these five coupling constants, 
\begin{subequations}
\label{eq:rel_of_cc}
\begin{align}
    \alpha_{PV}&=\frac{1}{3}(2\alpha_S+3\alpha_{tS}+2\alpha_{V}+3\alpha_{tV}+6\alpha_T)
    \label{eq:alpha_PV}\\
    \alpha_{tPV}&=\frac{1}{9}(2\alpha_S+3\alpha_{tS}+5\alpha_{V}-6\alpha_{tV}+6\alpha_T)
    \label{eq:alpha_tPV}\\
    \alpha_{tT}&=\frac{1}{18}(-\alpha_S+3\alpha_{tS}+2\alpha_{V}-6\alpha_{tV}+6\alpha_T).
    \label{eq:alpha_tT}
\end{align}
\end{subequations}
The PCF-PK1 model of Ref. \cite{PCF-PK1} also assumes a density dependence for 
$\alpha_S,~\alpha_V,~\alpha_{tS}$, and $\alpha_{tV}$ in a form of 
\begin{equation}
    \alpha_i(\rho_B)=\alpha_i(\rho_0)f_i(x),~~~i=S,V,tS, {\rm and}~tV, 
\end{equation}
where $\rho_B$ and $\rho_0$ are the baryon density and the nuclear saturation density, respectively, 
and $x$ is defined as $x\equiv \rho_B/\rho_0$. 
The function $f_i(x)$ is defined as 
\begin{equation}
    f_i(x)=a_i\,\frac{1+b_i(x+d_i)^2}{1+c_i(x+d_i)^2},
\end{equation}
where $a_i,b_i,c_i$, and $d_i$ are model parameters.

To construct the densities, 
one needs to determine the single-particle wave function $u(\bm{p},s,t)$ by 
solving the Dirac equation obtained by 
taking the variation of the the energy density $\mathcal{E}$ with respect to 
$u^{\dagger}(\bm{p},s,t)$, 
\begin{align}
\label{eq:Dirac_eq}
    \left[ \bm{p}\cdot\bm{\alpha}+M^*\beta-V_{PV}\Sigma_z+2V_T\beta\Sigma_z\right]u(\bm{p},s,t)\notag\\
    =\epsilon_{\bm{p},s,t}^*u(\bm{p},s,t).
\end{align}
In this equation, we have defined the effective mass, $M^*$, the effective energy, $\epsilon_{\bm{p},s,t}^*$, and the self 
energy for the $PV$ and $T$ channels, $V_{PV}$ and $V_T$, as
\begin{subequations}
 \begin{align}
    M^*&=M+\alpha_S\rho_S+\tau_3\alpha_{tS}\rho_{tS},\\
    \epsilon_{\bm{p},s,t}^*&=\epsilon_{\bm{p},s,t}-\alpha_V\rho_V-\tau_3\alpha_{tV}\rho_{tV},\\
    V_{PV}&=\alpha_{PV}\rho_{PV}+\tau_3\alpha_{tPV}\rho_{tPV},\\
    V_T&=\alpha_T\rho_T+\tau_3\alpha_{tT}\rho_{tT}.
\end{align}
\end{subequations}
The $4\times4$ matrix appearing on the left-hand side of Eq. \eqref{eq:Dirac_eq} reads 
\begin{align}
\label{eq:1p-ham}
    \left(
    \begin{array}{cc}
        M^*-(V_{PV}-2V_T)\sigma_z & \bm{p}\cdot\bm{\sigma} \\
        \bm{p}\cdot\bm{\sigma} & -M^*-(V_{PV}+2V_T)\sigma_z
    \end{array}
    \right).
\end{align}
When nuclear matter is not spin-polarized, the self-energies $V_{PV}$ and $V_{T}$ vanish, and 
one can solve the Dirac equation \eqref{eq:Dirac_eq} analytically for given densities. 
In contrast, for spin-polarized matter, it is much more difficult to obtain the analytic solution of the 
Dirac equation, and we thus numerically diagonalize the 4$\times$4 matrix, \eqref{eq:1p-ham}. 
The self-consistent solution for a given baryon density $\rho_B$ is obtained iteratively as follows.  
First, we solve the Dirac equation by setting $\rho_i=0$ for all $i$ except for $\rho_V$, which is set to be $\rho_V=\rho_B$,  
and then construct the density $\rho_i$ with the solutions of the the Dirac equation. 
This provides the initial value of the density for each coupling channel, $\rho_i$. 
We then numerically diagonalize 
the matrix \eqref{eq:1p-ham} for each momentum satisfying $|\bm{p}|<p_{k_F,s,t}$, where $p_{k_F,s,t}$ refers to the 
Fermi momentum evaluated in the Thomas-Fermi approximation for the nucleons with spin $s$ and isospin $t$.
For simplicity, we shall ignore the distortion of the Fermi surface \cite{Maruyama,Maruyama2018} induced by a spin polarization.
Notice that, since Eq. \eqref{eq:Dirac_eq} has the rotational symmetry around the $z$-axis, it is sufficient to consider the 2-dimensional momentum space $(p_{\perp},p_z)$, assuming $p_x=p_y=p_{\perp}$.
Second, we calculate each density, $\rho_i$, using the so obtained solution, $u(\bm{p},s,t)$. 
We repeat this procedure until 
the densities  $\rho_i$ are converged to certain values.


\section{Numerical Results}
\label{sec:results}

\begin{figure*}[t]
    \centering
    \includegraphics[width=0.8\linewidth]{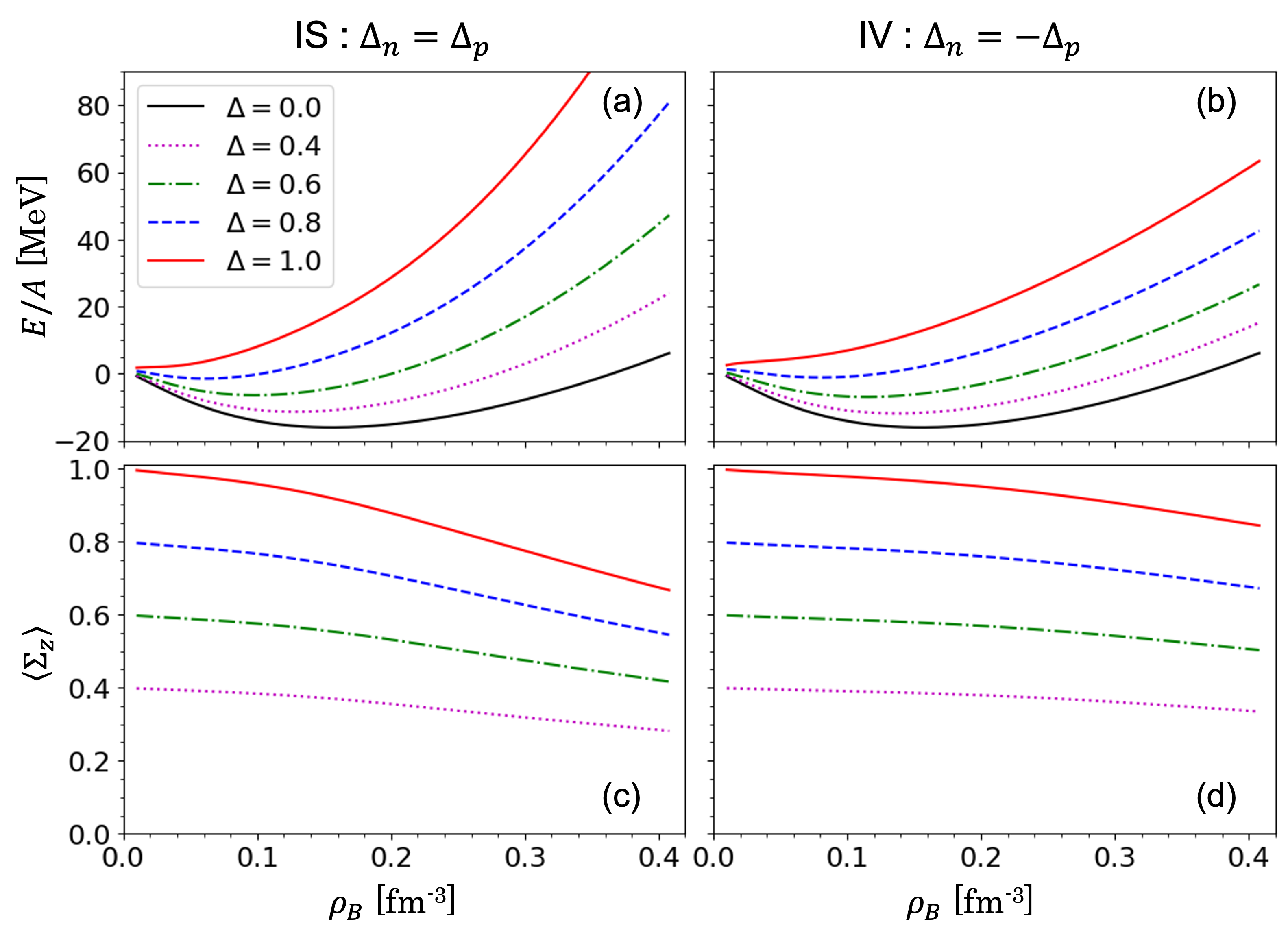}
    \caption{(a), (b): The EOS of spin-polarized symmetric nuclear matter 
    for (a) the isoscalar (IS) and (b) the isovector (IV) spin polarizations
    for various values of $\Delta$ defined by 
    Eq. \eqref{eq:Delta}. 
   (c), (d): The expectation value of the spin operator $\Sigma_z$ for the IS and the IV spin polarizations, respectively. }
    \label{fig:eos}
\end{figure*}

The EOS of spin-polarized matter is characterized by the baryon debsity, $\rho_B$, the isospin asymmetry, $\delta$, and the spin polarization rate $\Delta_t$ for nucleons with isospin $t$.
These are expressed as,
\begin{gather}
    \rho_B=\rho_V,\\
    \delta=\frac{\rho_n-\rho_p}{\rho_B},\\
    \label{eq:Delta}
    \Delta_t=\frac{\rho_{t,s=1}-\rho_{t,s=-1}}{\rho_t},
\end{gather}
where $\rho_{t,s}$ and $\rho_t~(t=n,p)$ are given by
\begin{align}
    \rho_{t,s}&=\int \frac{d^3p}{(2\pi)^3} u^\dagger(\bm{p},s,t)u(\bm{p},s,t), \\ 
    \rho_t&=\sum_{s}\rho_{t,s}. 
\end{align}
For simplicity, in this work we assume $|\Delta_n|=|\Delta_p|\equiv\Delta$.
We consider two cases: one is the isoscalar (IS) spin polarization, where 
the spins of neutrons and protons are polarized in the same direction, that is, 
$\Delta_n=\Delta_p$, while the other is the isovector (IV) spin polarization, where 
the spins of neutrons are polarized in the opposite direction to those of protons, that is, 
$\Delta_n=-\Delta_p$. 
Notice that only the IV polarization was considered in the previous study by Khoa et al. \cite{Khoa_2020,Khoa_2022}. 

Figs. \ref{fig:eos} (a) and (b) show the EOS of spin-polarized symmetric matter calculated by the relativistic point-coupling model.  
Fig. \ref{fig:eos} (a) shows the result of the IS polarization, while Fig. \ref{fig:eos} (b) shows that of the IV polarization.
One can see that 
the energy of nuclear matter for a given value of $\Delta$ is larger for the IS polarization than for the IV polarization.
This difference is due to the different contribution of the PV and T channels. 
That is, for the IS(IV) polarization, the isoscalar (isovector) components of these channels contribute. 
In any case, these contribution vanish in the Hartree approximation, indicating an importance 
of including the Fock terms.

It should be noted that the spin index, $s$, 
in the relativistic wave function $u(\bm{p},s,t)$ 
does not represent the spin along the $z$-axis, 
because the single-particle Hamiltonian in Eq. \eqref{eq:Dirac_eq} does not commute with the spin operator along the $z$-axis, $\Sigma_z$, except for $\bm{p}=0$, and thus the wave function $u(\bm{p},s,t)$ is not an eigenvector of $\Sigma_z$. 
That is, 
$\Delta$ 
should be regarded simply as a convenient parameter for numerical calculations, 
and a more consistent treatment is to calculate the expectation value of 
$\Sigma_z$.  
Notice that this problem is inherent to the relativistic approach, 
as the expectation value of $\Sigma_z$ coincides to $\Delta$ 
in the non-relativistic framework. 
Figs. \ref{fig:eos} (c) and (d) show the expectation value of $\Sigma_z$, $\ev{\Sigma_z}$. 
One can see that $\ev{\Sigma_z}$ is smaller than $\Delta$ especially in the high density region.
In comparison to the results of Khoa et al. \cite{Khoa_2020,Khoa_2022} for the IV polarization, 
we obtain a qualitatively similar EOS as a function of $\Delta$, even though 
our EOS is somewhat softer. 

The EOS of spin unpolarized matter is often expanded in the isospin asymmetry $\delta$ as
\begin{align}
\label{eq:expansion_of_eos}
    \frac{E}{A}(\rho_B,\delta) = \frac{E}{A}(\rho_B,\delta=0) + S(\rho_B)\delta^2 + \mathcal{O}(\delta^4),
\end{align}
in which $S(\rho_B)$ is the symmetry energy expressed as
\begin{align}
\label{eq:sym_ene}
    S(\rho_B) = \frac{1}{2}\left.\frac{\partial^2}{\partial\delta^2}\frac{E}{A}\right|_{\delta=0}.
\end{align}
The slope parameter $L$ is defined as 
\begin{align}
\label{eq:slope}
    L = 3\rho_0\left.\frac{\partial S}{\partial\rho_B}\right|_{\rho_B=\rho_0}. 
\end{align}
It has been well recognized that 
the 4th-order or the higher-order terms with respect to $\delta$ in Eq. \eqref{eq:expansion_of_eos} 
are negligibly small \cite{Khoa_1996,Togashi_2013}. 
In other words, one can approximately obtain the symmetry energy $S(\rho_B)$ as 
\begin{align}
\label{eq:sym_ene_approx}
    S(\rho_B) \simeq \frac{E/A(\rho_B,\delta)-E/A(\rho_B,\delta=0)}{\delta^2}, 
\end{align}
instead of Eq. \eqref{eq:sym_ene}. 
By analogy with the symmetry energy $S(\rho_B)$, one can 
introduce the spin symmetry energy $W(\rho_B,\delta)$ \cite{Khoa_2022}. 
In the relativistic framework, the original expression for $W(\rho_B,\delta)$ in Ref. \cite{Khoa_2022}
can be modified as 
\begin{align}
\label{eq:spin_sym_ene}
    W(\rho_B,\delta) = \frac{E/A(\rho_B,\delta,\Delta)-E/A(\rho_B,\delta,\Delta=0)}{\left[\ev{\Sigma_z}(\rho_B,\delta,\Delta)\right]^2}.
\end{align}
The definition of the spin symmetry energy $W(\rho_B,\delta)$ differs from that of Ref. \cite{Khoa_2022} 
in that the numerator of $W(\rho_B,\delta)$ is not $\Delta$ but $\ev{\Sigma_z}$.
From Eq. \eqref{eq:spin_sym_ene}, one can define the spin slope parameter $L_s$ as \cite{Khoa_2022}
\begin{align}
\label{eq:spin_slope}
    L_s = 3\rho_0\left.\frac{\partial W}{\partial\rho_B}\right|_{\rho_B=\rho_0}.
\end{align}

In our calculations, we obtain the slope parameter of $L=79.9$ MeV for spin unpolarized matter. 
This value is consistent with the empirical value $L=58.7\pm28.1$ MeV \cite{Oertel}.
The spin slope parameters in our calculations read $L_s=151.6$ MeV for the IS polarization and $L_s=73.1$ MeV for the 
IV polarization.
In the non-relativistic calculation by Khoa et al. \cite{Khoa_2022}, they obtained the spin slope parameter for the 
IV polarization to be $L_s\simeq96$ MeV.
This value is somewhat larger than the value in our calculations, 
that is, their EOS for IV spin polarized matter is stiffer than ours.


\section{correlation between the slope parameters}
\label{sec:correlation}

In order to investigate a possible probe to constrain the spin slope parameter, $L_s$, we first study 
the correlation between the slope parameter $L$ and the spin slope parameter $L_s$.
Since the measurement of the neutron skin thickness of finite nuclei can constrain the slope parameter $L$ \cite{Roca-maza}, 
the spin slope parameter can also be constrained if 
the spin slope parameter $L_s$ is significantly correlated with the slope parameter $L$.  
For this purpose, 
we apply a similar method as that proposed in Ref. \cite{Inakura}, in which 
the slope parameter is varied by changing the density dependence of the coupling constants. 
To this end, 
we introduce a new parameter $y$ and modify the coupling constant for the isovector-vector channel, $\alpha_{tV}(\rho_B)$, as
\begin{align}
    \alpha_{tV}(\rho_B) \rightarrow (1-y) \alpha_{tV}(\rho_B) + y \alpha_{tV}(\rho_0).
    \label{eq:alpha_tV}
\end{align}
Notice that this modification keeps 
the value of $\alpha_{tV}(\rho_B)$ at the saturation density, $\rho_0$. 
It therefore modifies 
only the slope of the EOS while the EOS at the saturation density is unaffected. 
Notice that one could also modify the coupling constant for the isovector-scalar channel, $\alpha_{tS}(\rho_B)$, instead of 
$\alpha_{tV}$, but we have confirmed that the results do not significantly change and thus we 
only show the results with the modification in $\alpha_{tV}$.

\begin{figure}[t]
    \centering
    \includegraphics[scale=0.36]{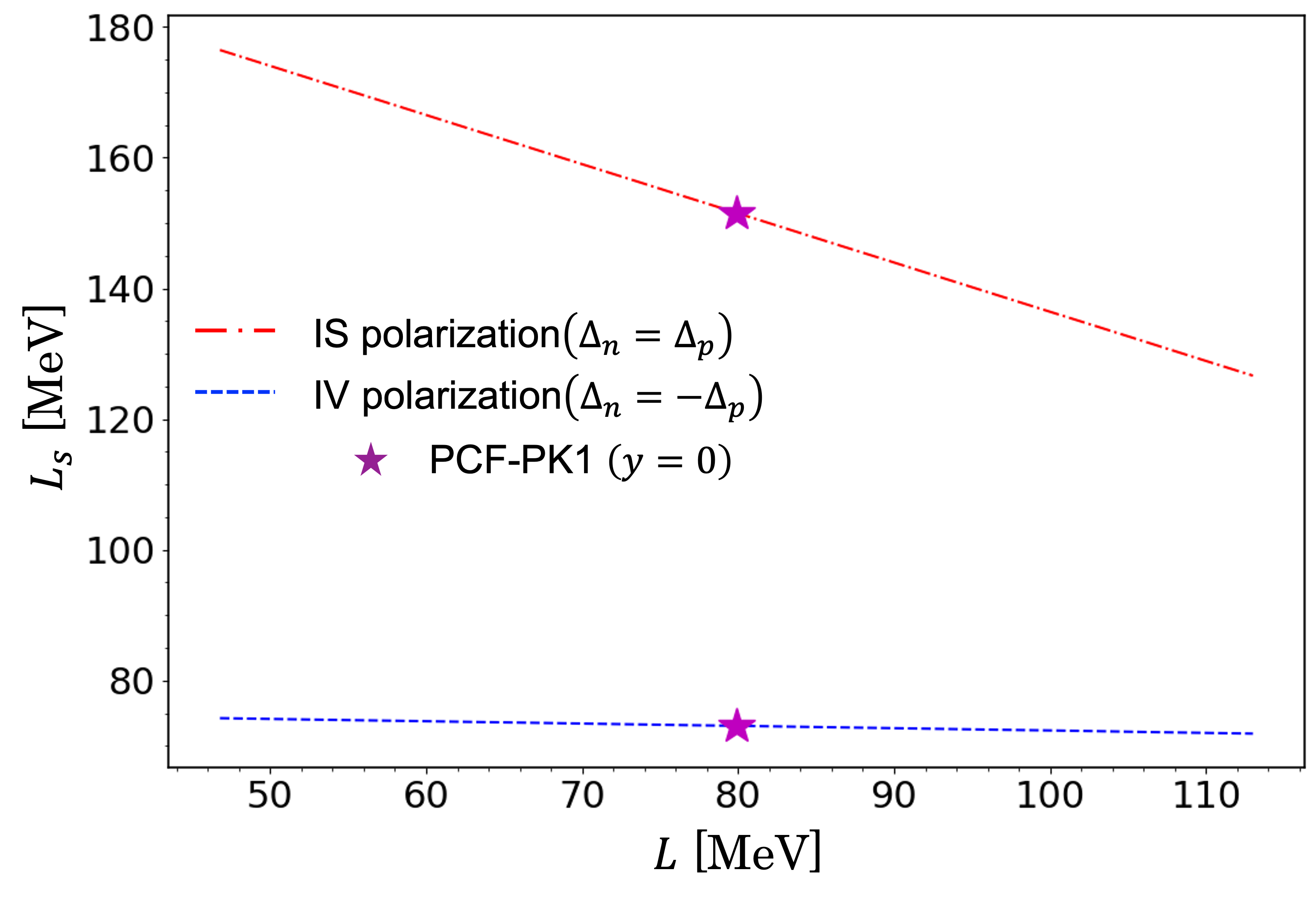}
    \caption{The correlation between the slope parameter, $L$, and the spin slope parameter, $L_s$. The red dot-dashed line represents the spin slope parameter for the IS polarization, while the blue dashed line represents that for the IV polarization. 
    The stars show the result with the original PCF-PK1 parameter set, i.e., with $y=0$.}
        \label{fig:slope_corr}
\end{figure}

Fig. \ref{fig:slope_corr} shows the correlation between the slope parameter $L$ and the spin slope parameter $L_s$. 
The values of the slope parameters obtained with the original value of the coupling constant $\alpha_{tV}$, 
that is, $y=0$, are shown by the stars. 
In the case of the IS spin polarization shown by the red dot-dashed line, 
the slope parameter $L$ and the spin slope parameter $L_s$ exhibit a negative linear correlation.
On the other hand, in the case of the IV spin polarization shown by the 
blue dashed line, the spin slope parameter $L_s$ is nearly independent of the slope parameter $L$ and remains almost a constant. 
This indicates that the spin slope parameter for the IS spin polarization 
can be constrained by measuring the neutron skin thickness, whereas that for the IV polarization 
cannot be constrained in this way.

\begin{figure}[t]
    \centering
    \includegraphics[scale=0.36]{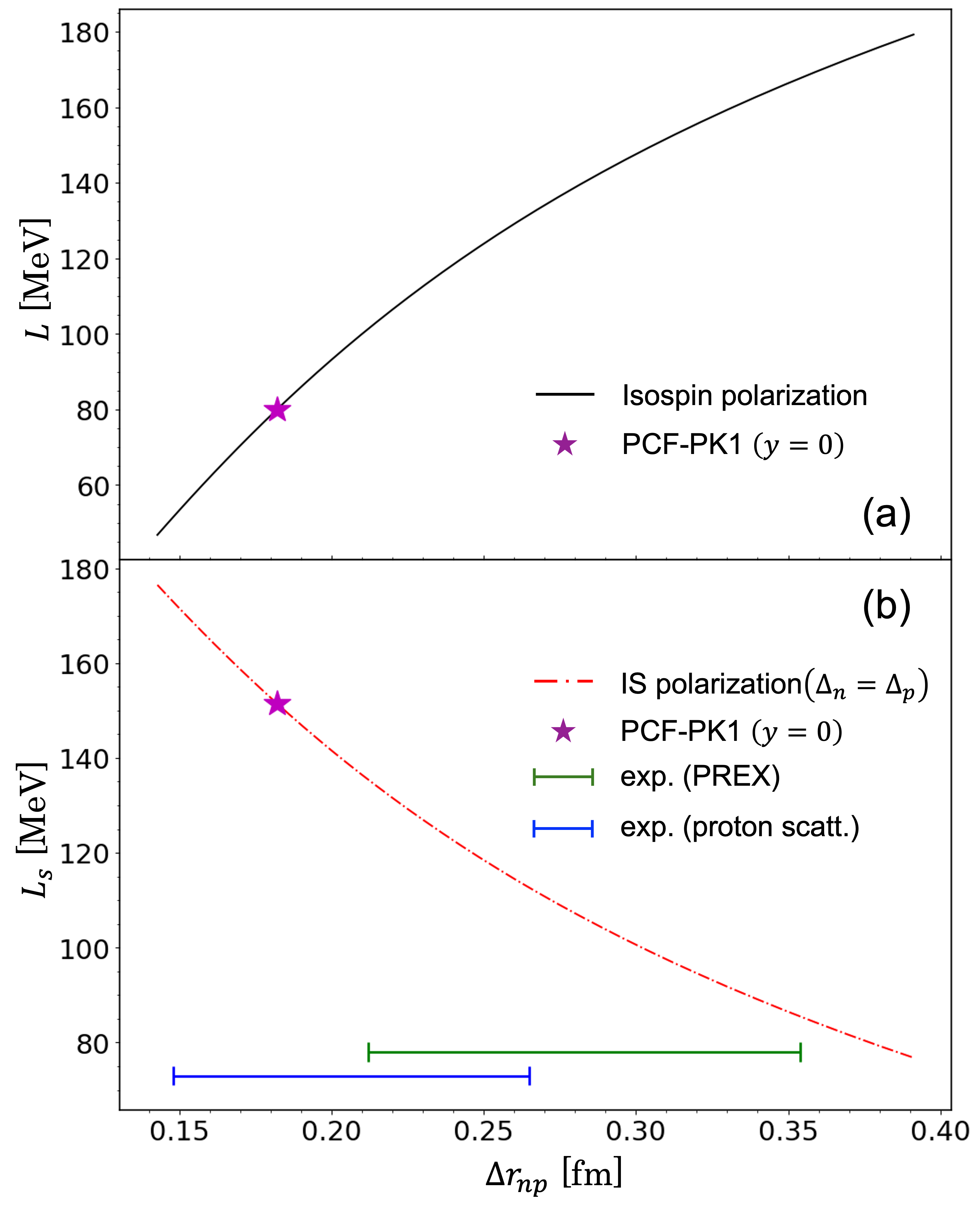}
    \caption{Same as Fig. \ref{fig:slope_corr}, but for 
    the correlation between the slope parameter $L$ and the neutron skin thickness of \ce{^{208}Pb} (the upper panel)  
    and 
    the correlation between the spin slope parameter $L_s$ for the IS polarization and the neutron skin thickness of \ce{^{208}Pb} (the lower panel). The green (blue) range denotes the neutron skin thickness from the PREX (the proton elastic scattering) experiment.}
    \label{fig:slope_corr2}
\end{figure}

Figs. \ref{fig:slope_corr2} (a) and \ref{fig:slope_corr2} (b) show  
the correlation of the neutron skin thickness $\Delta r_{np}$ of \ce{^{208}Pb} with 
the slope parameter $L$ and the spin slope parameter $L_s$ for the IS polarization, respectively. 
One can see that $L$ and $\Delta r_{np}$ have a positive correlation, as has been known well \cite{Roca-maza}. 
On the other hand, 
one can find a negative correlation between $L_s$ for the IS polarization and the neutron 
skin thickness, originated from the negative correlation between $L$ and $L_s$ shown in Fig.  \ref{fig:slope_corr}. 
From the experimental value of the neutron 
skin thickness of \ce{^{208}Pb}, 
the spin slope parameter for the IS polarization is 
estimated to be 86 MeV $\lesssim L_s\lesssim$ 135 MeV and 
113 MeV $\lesssim L_s\lesssim$ 173 MeV from the 
PREX \cite{PREX} and the proton elastic scattering \cite{Zenihiro} measurements, respectively.
Those results are summarized in Tab. \ref{tab:L_and_Ls}, together with the constrained value of $L$. 
Notice that 
the spin slope parameter for the IV polarization remains almost a constant as a function of 
the neutron skin thickness of \ce{^{208}Pb} and cannot be constrained from the skin thickness. 

The correlation between the slope parameters shown in Fig. \ref{fig:slope_corr} 
reflects the relation among the coupling constants shown in  Eq. \eqref{eq:rel_of_cc}. 
According to Eq. \eqref{eq:sym_ene_approx} for the definition of the symmetry energy, the symmetry energy can be obtained from 
the EOS of symmetric matter and neutron matter.
For symmetric matter, only $\rho_S$ and $\rho_V$ have non-zero values in Eq. \eqref{eq:energy_density} while 
the other densities vanish. On the other hand, for neutron matter, $\rho_{tS}$ and $\rho_{tV}$, 
in addition to $\rho_S$ and $\rho_V$, are non-zero and the others vanish.
Since, for neutron matter, $\rho_{tS}$ and $\rho_{tV}$ are approximately the same as $\rho_B$ at the saturation density, 
the symmetry energy $S(\rho_B)$ is approximately calculated as 
\begin{align}
    S(\rho_B) & \simeq \frac{E}{A}(\rho_B,\delta=1,\Delta=0)
    - \frac{E}{A}(\rho_B,\delta=0,\Delta=0) \notag \\
    & = \Delta E_{\text{kin}} + \frac{1}{2\rho_B} (\alpha_{tS}\rho_{tS}^2 + \alpha_{tV}\rho_{tV}^2) \notag \\
    & \simeq \Delta E_{\text{kin}} + \frac{1}{2} (\alpha_{tS} + \alpha_{tV} ) \rho_B,
\end{align}
where $\Delta E_{\text{kin}}$ represents the difference between the kinetic energy for symmetric matter and that for  
neutron matter.
The slope parameter $L$ can be calculated by differentiating the symmetry energy $S(\rho_B)$ with respect to the 
baryon density $\rho_B$ as, 
\begin{align}
\label{eq:slope_aapprox}
    L = L_{\text{kin}}^{(n)} + \frac{3}{2} (\alpha_{tS} + \alpha_{tV}) \rho_0,
\end{align}
where $L_{\text{kin}}^{(n)}$ is the contribution to the slope parameter from $\Delta E_{\text{kin}}$.
On the other hand, the spin symmetry energy for the IS polarization is obtained from the EOS of  
unpolarized and fully polarized symmetric matter.
For fully IS-polarized symmetric matter, $\rho_{PV}$ and $\rho_T$ as well as $\rho_S$ and $\rho_V$ are finite 
and the others vanish. 
Following a similar procedure as in $L$, 
the spin slope parameter can thus be approximately obtained as 
\begin{align}
\label{eq:IS_slope_approx}
    L_s^{\text{(IS)}} \simeq L_{s,\text{kin}}^{\text{(IS)}} + \frac{3}{2} ( -\alpha_{PV} + 2\alpha_T) \rho_0.  
\end{align}
Similarly, the spin slope parameter for the IV polarization is obtained as 
\begin{align}
\label{eq:IV_slope_approx}
    L_s^{\text{(IV)}} \simeq L_{s,\text{kin}}^{\text{(IV)}} + \frac{3}{2} (-\alpha_{tPV} + 2\alpha_{tT}) \rho_0,
\end{align}
in which 
we have used 
the fact that only $\rho_{tPV}$ and $\rho_{tT}$ as well as $\rho_S$ and $\rho_V$ have non-zero values 
for fully IV-polarized matter.

\begin{figure}[tb]
    \centering
    \includegraphics[width=\linewidth]{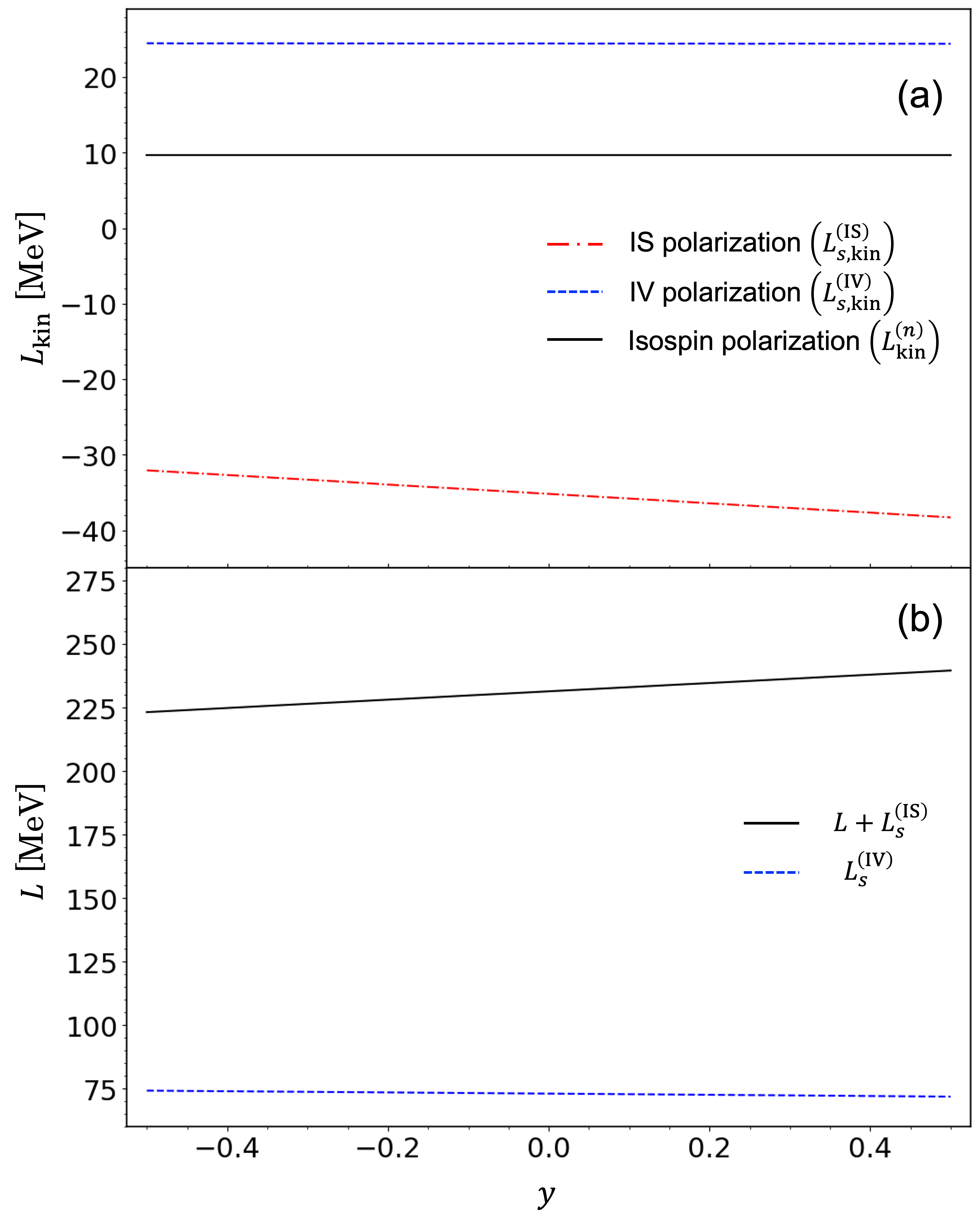}
    \caption{(a) The contributions of the kinetic energy shift to the slope parameter (the black solid line) 
    and the spin slope parameter (the red dot-dashed and the blue dashed lines for the IS and the IV polarizations, respectively). 
    (b) The dependence of the slope parameter and the spin slope parameter on $y$ defined by Eq. (\ref{eq:alpha_tV}). The black solid line represents the sum of the slope parameter and the spin slope parameter for the IS polarization, while the blue dashed line represents the spin slope parameter of IV spin-polarized symmetric matter.}
    \label{fig:kin_slope}
\end{figure}

In this way, 
the three slope parameters, $L$, $L_s^{\text{(IS)}}$ and $L_s^{\text{(IV)}}$, are expressed 
in terms of the kinetic terms and the potential terms with a linear combination of the coupling constants.
Because not all of the coupling constants are independent 
but are related through Eq. \eqref{eq:rel_of_cc}, 
$L$, $L_s^{\text{(IS)}}$ and $L_s^{\text{(IV)}}$ can be transformed into a simpler form.
Adding both sides of Eq. \eqref{eq:slope_aapprox} and Eq. \eqref{eq:IS_slope_approx} 
leads to 
\begin{align}
    L + L_s^{\text{(IS)}} \simeq L_{\text{kin}}^{(n)} + L_{s,\text{kin}}^{\text{(IS)}} - (\alpha_S+\alpha_V)\rho_0, 
\end{align}
with the expression for $\alpha_{PV}$ shown in Eq. \eqref{eq:alpha_PV}. 
The third term in this equation includes only $\alpha_S$ and $\alpha_V$, 
which are independent from $\alpha_{tV}$. 
Namely, the third term remains a constant even if the value of $y$ changes according to Eq. (\ref{eq:alpha_tV}).  
The first and the second terms are contributions from the kinetic energy and their behavior is non-trivial when $y$ is varied. 
However, numerical calculations 
indicate that $L_{\text{kin}}^{(n)}$ and $L_{\text{kin}}^{\text{(IS)}}$ remain nearly constants as a function of $y$, 
as is shown in 
Fig. \ref{fig:kin_slope} (a). 
That is, $L+L_s^{\text{(IS)}}\simeq const.$ as shown in Fig. \ref{fig:kin_slope} (b), 
and thus 
\begin{align}
    L \sim -L_s^{\text{(IS)}} + const..
\end{align}
This is consistent with 
the negative linear correlation between the slope parameter $L$ and the spin slope parameter $L_s^{\text{(IS)}}$ for the 
IS polarization shown in Fig. \ref{fig:slope_corr}. 

For the IV polarization, 
by substituting Eqs. \eqref{eq:alpha_tPV} and \eqref{eq:alpha_tT} into 
Eq. \eqref{eq:IV_slope_approx} one obtains 
\begin{align}
    L_s^{\text{(IV)}} \simeq L_{s,\text{kin}}^{\text{(IV)}} - \frac{1}{2} (\alpha_S+\alpha_V) \rho_0.
\end{align}
In this equation, 
the third term does not include the coupling constants of the isovector channels and remains a constant independently of $y$.
The first term is almost a constant as a function of $y$ as shown 
in Fig. \ref{fig:kin_slope} (a). 
Therefore, the spin slope parameter $L_s^{\text{(IV)}}$ for the IV polarization is independent of the slope 
parameter $L$, 
as shown in Fig. \ref{fig:kin_slope} (b).

Notice that the relations among the coupling constants \eqref{eq:rel_of_cc} are determined by the 
matrix of the Fierz transformation, and therefore, the correlation between the slope parameters stems 
from the relativistic structure of the point-coupling model.
In the above discussions, for simplicity we have omitted the derivative terms of 
the coupling constants originated from their density dependence. 
Those terms, however, do not essentially affect the conclusions as long as they are independent of $\alpha_{tV}$.

\begin{table}[t]
\centering
\caption{
The slope parameter $L$ and the spin slope parameter for the IS polarization $L_s^{\text{(IS)}}$ 
extracted from 
the experimental value of the neutron skin thickness of \ce{^{208}Pb}, $\Delta r_{np}$, 
from the PREX measurement \cite{PREX} and the proton elastic scattering \cite{Zenihiro}. 
}
\begin{tabular}{c|cc}
\hline \hline
    & PREX \cite{PREX} & p-scatt. \cite{Zenihiro} \\ \hline
   $\Delta r_{np}$ [fm] & $0.283\pm0.071$ & $0.211^{+0.054}_{-0.063}$ \\
   $L$ [MeV] & $101-168$ & $52-132$ \\
   $L_s^{\text{(IS)}}$ [MeV] & $86-135$ & $113-173$ \\ 
   \hline \hline
\end{tabular}
\label{tab:L_and_Ls}
\end{table}


\section{SUMMARY}
\label{sec:summary}

We have calculated the EOS of spin-polarized matter in the relativistic point-coupling model.
In this study, we have considered two types of spin polarization: one is the IS polarization and the other is the IV polarization.
For both of these, 
we have calculated the spin slope parameter in a similar way to the slope parameter 
for spin-unpolarized matter. 
To propose a probe to constrain the EOS of spin-polarized matter, 
we have investigated the correlation between the slope parameter and the spin slope parameter.
We have found a negative linear correlation between the slope parameter and the spin slope parameter in the case of the IS polarization, 
whereas for the IV polarization, the spin slope parameter was found to be nearly independent of the slope parameter and remain 
almost a constant.
We have shown that these behaviors of the spin slope parameters can be understood in terms of 
the relativistic structure of the model which we employed in this work.

Our finding suggests that the spin slope parameter for the IS polarization can be constrained by measuring the neutron skin 
thickness of finite nuclei. Using  
the experimental value of the neutron 
skin thickness of \ce{^{208}Pb}, 
we have extracted 
the spin slope parameter for the IS polarization of 
85 MeV $\lesssim L_s\lesssim$ 135 MeV and 
113 MeV $\lesssim L_s\lesssim$ 172 MeV from the 
PREX \cite{PREX} and the proton elastic scattering \cite{Zenihiro} measurements, respectively.

In this study, we have considered only symmetric nuclear matter.
It would be straightforward to extend the present work to 
the EOS for spin polarized matter with an arbitrary isospin asymmetry $\delta$.
In addition, although we have investigated in this paper the correlation between the slope parameter and the spin slope parameter, 
it would be more desirable to propose an observable for finite nuclei which is directly correlated with the spin 
slope parameter. Moreover,   
it is also important to constrain the $W(\rho_0,\delta)$ itself, that is, the spin symmetry energy at the saturation density.
We leave these as interesting future works.


\section*{acknowledgments}
We thank N. Hinohara, G. Col\`{o}, M. Cheoun, H. Togashi, K. Uzawa, and P.W. Zhao for useful discussions. 
This work was supported by the JSPS KAKENHI (Grant Nos.
JP23K03414, JP23H05434, JP25H01269, JP25H00402 and JP25K07322),
JST ERATO (Grant No. JPMJER2304),
JST SPRING (Grant No. JPMJSP2110),
the National Natural Science Foundation of China (Grant No. 12275359),
the Continuous Basic Scientific Research Project,
the Young Talent Cultivation Foundation of CIAE (Grant No. YC010270525794).

\bibliographystyle{apsrev4-2.bst}
\bibliography{main}
\end{document}